\documentclass[aps,twocolumn,amsmath,amssymb,preprintnumbers]{revtex4}
\usepackage{amsmath} \usepackage{amsfonts} \usepackage{amssymb}
\usepackage{bbm}
\usepackage{epsfig}
\usepackage{graphics}
\usepackage{graphicx}

\newcommand{\be}{\begin{equation}}
\newcommand{\ee}{\end{equation}}
\newcommand{\ba}{\begin{eqnarray}}
\newcommand{\ea}{\end{eqnarray}}

\begin{document}

\title[ ]{Counting similarities between tetraquark and mesonic/gluonic operators}

\author{ Ming Chen}
\email{saintiming@emails.bjut.edu.cn}
\author{Yong-Chang Huang}
\email{ychuang@bjut.edu.cn}
\affiliation{Institute of Theoretical Physics, Beijing University of Technology, Beijing 100124, China}

\date{\today $\vphantom{\bigg|_{\bigg|}^|}$}

\begin{abstract}
After the study of the preclusion of exotic meson states in large-$N_c$ limit QCD, combining Weinberg's opposite proposal, we get different counting orders for a tetraquark operator to create or destroy an one-tetraquark state. Meanwhile, by comparing tetraquark operator with the mesonic and gluonic operators, we find that tetraquark operators are similar with mesonic and gluonic operators in the counting. Furthermore, we find a mixing of different kinds of operators.
\vspace{1em}

keywords: large-$N_c$ QCD; tetraquark                       \hfill                        PACS numbers: 11.15.Pg, 12.38.Lg, 13.25.Jx, 14.40.Rt
\end{abstract}

\maketitle

\section{Introduction}
As Witten argued \cite{witten}, large-$N_c$ limit QCD($\mathbf{F}$) \cite{thooft} could lead to the remarkable preclusion of exotic meson states. This preclusion was explicitly discussed in Coleman's lecture \cite{coleman}.\\

The connected correlators of physical, local or composite operators $\mathcal{O}_n$ which involve at most one trace over color indices, can be written as:
\begin{equation}
\big<\!\mathcal{O}_1(x_1) \!\cdots\! \mathcal{O}_n(x_n)\!\big>_c\!\!\!\!=\!(iN_c)^{-n} \!\!\! \frac{\delta}{\delta J_1(x_1)}\! \cdots \!\frac{\delta}{\delta J_n(x_n)} ln\mathcal{Z}_J |_{J=0},
\end{equation}

where $\mathcal{Z}_J$ is the generating functional of the general $SU(N_c)$ non-Abelian vector gauge theory with an operator source term which reads \cite{reports}:
\begin{equation}
\mathcal{Z}_J=\!\!\int\!\! \mathcal{D}A \mathcal{D} \bar{\psi} \mathcal{D} \psi exp \{ iN_c \!\!\int \!\!dtd^3x (\tilde{\mathcal{L}}[A,\psi,\bar{\psi}]+J_n(x)\mathcal{O}_n(x)) \}.
\end{equation}

For our interest, we will currently focus on the color-singlet quark bilinears \cite{weinberg2}:
\begin{equation}
\mathcal{B}(x)=\sum_{a=1}^{N_c} \bar{q^a}(x) \Gamma_i q^a(x) - \sum_{a=1}^{N_c} \big<\bar{q^a}(x) \Gamma_i q^a(x)\big>_0.
\end{equation}

Here, $q^a$ denotes the canonically normalized quark fields, with $a$ a $N_c$-component $SU(N_c)$ color index, summing over all $N_c$ colors. Meanwhile, with spin and flavor indices suppressed, $\Gamma_i$ are $N_c$-independent matrices containing spin and flavor information. $\big< \cdots \big>_0$ is the vacuum expectation value. Because it is not necessary in our following arguments, we will just drop it and rewrite the bilinears as \cite{lebed}:
\begin{equation}
\mathcal{B}_i (x)=\sum_{a=1}^{N_c} \bar{q^a}(x) \Gamma_i q^a(x).
\end{equation}

From Eq.(1), in the case of mesons, we can easily get meson interactions from connected correlators of the form in the following:
\begin{equation}
<\mathcal{B}_1(x_1) \cdots \mathcal{B}_n(x_n)>_c.
\end{equation}

In this case, the dominant diagrams contributing to the correlators are just the planar ones with a single quark loop which runs at the edge of them, and they will give an extra counting $N_c$ \cite{witten}. As a result, Eq.(5) counts like $O(N_c^{1-n})$.\\

If we further consider the normalization of the two-point correlator, the order for a mesonic operator $\mathcal{B}$ to create or destroy a meson state is actually $O(\sqrt{N_c})$. This changes the counting of Eq.(5) to $O(N_c^{1-n/2})$. Because we will deal with decay width later and we should use its related \textbf{LSZ} reduction formula, this normalization is necessary.\\

Complying with the above logic, we first discuss the preclusion of tetraquark states and the problem of it in Sec.II. Then in Sec.III, we get the counting similarities between tetraquark operator and mesonic/gluonic operators. Sec.IV is devoted to some discussions, while Sec.V gives the summary and conclusions.

\section{Preclusion of tetraquark}
Considering the color-singlet operator formed by two quark fields and two antiquark fields of the form like this:
\begin{equation}
\mathcal{Q}(x)=\sum_{ij} C_{ij} \mathcal{B}_i(x) \mathcal{B}_j(x),
\end{equation}

where $C_{ij}=C_{ji}$ are some $N_c$-independent numerical coefficients describing the correlation of bilinears at spacetime point $x$.\\

To show exactly the nonexistence of exotic meson states, Coleman studied $\mathcal{Q}$'s two-point function \cite{coleman}:
\begin{eqnarray}
<\mathcal{Q}^{\dagger}(x)\mathcal{Q}(y)> \!\!&=&\!\! \sum_{ijkl} C_{ij} C^{\ast}_{kl}\big[\big<\mathcal{B}_i^{\dagger}(x) \mathcal{B}_k(y)\big> \big<\mathcal{B}^{\dagger}_k(x)\mathcal{B}_l(y)\big> \nonumber\\
                                               & & + \big<\mathcal{B}^{\dagger}_i(x) \mathcal{B}^{\dagger}_j(x) \mathcal{B}_k(y) \mathcal{B}_l(y)\big>_c \big].
\end{eqnarray}

By using the counting of Eq.(5), we can see from the right-hand side of Eq.(7) that, the first disconnected term is of  $O(N_c^0)$, while the second connected term is of  $O(1/N_c)$.\\

At the same time, the one-tetraquark pole can only appear in the later connected term. It is clearly that any one-tetraquark state will be at least suppressed by a factor $1/N_c$. Therefore, the two-point function of Eq.(7) can only create two-meson states.\\

This conclusion, i.e., the preclusion of tetraquark states, is consistent with Witten's argument \cite{witten}, and it has been accepted for decades.\\

However, as one can see, in our previous argument, what we concerned about are just the correlation functions of quark bilinears, without the consideration of their scattering amplitudes.\\

As Weinberg argued \cite{weinberg2}, if we simply drop the second connected term in the right-hand side of Eq.(7) based on the general suppressed factor $1/N_c$, then we will lose any possibility for the ordinary mesons scattering in large-$N_c$ limit.\\

Moreover, Coleman's conclusion actually implies that, with tetraquark mass be independent of $N_c$, the decay width must grow as some positive power of $N_c$. Then, the lifetime of the supposed tetraquark meson will be too short for it to be observed as a distinct particle. However, this implication is farfetched because nothing can guarantee it.\\

One can check the above argument through dealing with the decay width. In Eq.(7), if there is an one-tetraquark pole (propagator) in the leading part of the connected term which is of $O(1/N_c)$, then after the normalization of this two-point correlator, the operator $\mathcal{Q}$ creates or destroys a tetraquark state will be of $O(\sqrt{N_c})$.\\

If we further consider the connected three-point function, which shows the decay of such a tetraquark into two ordinary mesons of type $\alpha$ and $\beta$:
\begin{eqnarray}
\big<\mathcal{Q}(x)^{\dagger}\mathcal{B}_{\alpha}(y)\mathcal{B}_{\beta}(z)\big>\!\! &=& \!\!
 \sum_{ij} C_{ij} \big<\mathcal{B}_i(x)^{\dagger} \mathcal{B}_{\alpha}(y)\big> \big<\mathcal{B}_j(x)^{\dagger}\mathcal{B}_{\beta}(z)\big> \nonumber\\
                                              &+& \big<\mathcal{Q}(x)^{\dagger} \mathcal{B}_{\alpha}(y) \mathcal{B}_{\beta}(z)\big>_c,
\end{eqnarray}

we can get the counting of the first disconnected term in its right-hand side, which is of $O(N_c^0)$. For its second connected term, by using the counting of Eq.(5), it counts like $O(1/\sqrt{N_c})$, corresponding to the trilinear vertex (pole) which gives a decay width of the tetraquark decaying into two ordinary mesons. Such a decay width is proportional to $1/N_c$, instead of any positive power of $N_c$, just like the decay width of ordinary mesons \cite{witten}.\\

As a result, the subleading connected term of Eq.(7) should not be hastily dropped, it may contain distinct tetraquarks. This denies the preclusion of tetraquarks and opens a door to the existence of them.

\section{Counting similarities}
Along with Weinberg's conclusion, one should keep in mind that, in order to get what we want, we used an assertion that an one-tetraquark pole should exist in the leading part of the connected term of Eq.(7), and it should be of $O(1/N_c)$.\\

Then the operator $\mathcal{Q}$ creates or destroys a tetraquark meson state, which should be of $O(\sqrt{N_c})$, just like mesonic operator $\mathcal{B}$ in the counting. However, when we see into the subtle flavor structure of $\mathcal{Q}$, the result will be complicated.\\

As Witten argued \cite{witten2}, the counting of correlators' dominant diagrams originates from the color degrees of freedom (DOF) and vertices coupling. Because a tetraquark operator contains two quarks and two anti-quarks, the ways how they contract will be crucial for color DOF and consequently affect the counting of the correlators.\\

From Peris' work \cite{peris}\cite{commu2}, its ``Type-A'' tetraquark operator, which takes the form $\bar{q}_A q_B \bar{q}_B q_D$ and with the flavor index `$_B$' contracted, will have the same flavor representation as ordinary $\bar{q}_A q_D$ mesons.\\

It is similar with an ordinary mesonic operator because they have the same DOF. Therefore, the order ($O(\sqrt{N_c})$) for such a $\mathcal{Q}$ to create or destroy a tetraquark state is the same as Weinberg's. In the meantime, the consequent decay width ($O(1/N_c)$) is just like that of the ordinary mesons \cite{witten}.\\

In fact, with such understanding, if we just want to get the aimed decay width from the practical purpose, we can replace $\mathcal{Q}$ with $\mathcal{B}$ in the connected term of Eq.(8), this will make no difference in the counting. One can check it from the counting of Eq.(5) ($O(N_c^{1-n/2}$)), with $n=3$, we can get the same decay width ($O(1/N_c)$). This reveals a similarity between tetraquarks and mesons in the counting.\\

For ``Type-B'' tetraquark operator \cite{peris}\cite{commu2}, it actually contains two quark loops and corresponds to two separated mesons. As a result, it cannot contain any tetraquark.\\

In order to make this kind of diagram connected, we should add at least two gluons between these two quark loops. There will be four more three-point vertices which generate four coupling constants, and they will contribute a factor $1/N_c^2$. This will turn down the overall counting to $O(N_c^0)$. This equally means that the order for $\mathcal{Q}$ to create or destroy a tetraquark state is $O(N_c^0)$, because the two-point tetraquark correlation $\big<\mathcal{Q}\mathcal{Q}\big>$ is of order $O(N_c^0)$, which has already been normalized automatically.\\

This reminds us of the gluons. Let $\mathcal{G}_m$ be a purely gluonic, gauge-invariant and Hermitian operator, with the appropriate quantum numbers to describe a glueball. The form of gluon interactions is similar with Eq.(5) ($<\mathcal{G}_1 \dots \mathcal{G}_m>_c$), while now the counting is of $O(N_c^{2-m})$ because the dominant diagrams give an extra counting $O(N_c^2)$ \cite{witten}.\\

Therefore, the two-point glueball correlation $\big< \mathcal{G}_1\mathcal{G}_2 \big>$ is of $O(N_c^0)$. We can then obtain the operator $\mathcal{G}$ to create or destroy a glueball state which is of $O(N_c^0)$. It is exactly the same as $\mathcal{Q}$. As a result, for ``Type-B'' tetraquark, it also has a similarity with the ordinary glueball in the counting.\\

Furthermore, glueball-meson interactions and the mixing processes are described by correlation of the form:
\begin{equation}
<\mathcal{G}_1 \cdots \mathcal{G}_m \mathcal{B}_1 \cdots \mathcal{B}_n >_c,
\end{equation}
which counts like $O(N_c^{1-m-n/2})$ \cite{witten}\cite{reports}.\\

If we replace the gluonic operator $\mathcal{G}$ with the tetraquark operator $\mathcal{Q}$, with $m=1,n=2$, we can easily check from the subleading part of Eq.(8) that the vertex for a tetraquark decaying into two ordinary mesons will be of $O(1/N_c)$. This will lead to a decay width of $O(1/N_c^2)$, the same as ``Type-B''. This approves that ``Type-B'' tetraquark operator indeed has a similarity with the ordinary gluonic operator in the counting.\\

We have already proved that ``Type-A'' tetraquark operator is similar with ordinary mesonic operator, and now we verifies that ``Type-B'' tetraquark operator is similar with the ordinary gluonic operator. In conclusion, although tetraquarks and mesons/glueballs are different from many points, their operators seem very similar in the counting, especially for the calculation of their corresponding decay widths.

\section{Discussions}
From Sec.III, ``Type-B'' tetraquark has some contradiction with Weinberg's result. Its corresponding decay width is of $O(1/N_c^2)$, narrower than ``Type-A's'' or Weinberg's, which are all of $O(1/N_c)$.\\

We can see that such contradiction originates from different contracting ways of quarks contained in a tetraquark operator. In fact, these different ways attribute to the subtle local and nonlocal properties of a tetraquark operator. As Lebed argued \cite{lebed}\cite{commu}, the local and nonlocal properties of a tetraquark operator are very important to the tetraquark states which only appear in the local situation.\\

Complying with this point, with one pair of quark-antiquark contracted, ``Type-A'' tetraquark operator actually behaves more like an ordinary mesonic operator than a localized tetraquark operator; while for ``Type-B'' tetraquark operator, without contracted quarks and with the added gluons, it is indeed a localized tetraquark operator and its order to create or destroy an one-tetraquark state is of $O(N_c^0)$.\\

In order to maintain Weinberg's result, Lebed \cite{lebed} starts from the nonlocal situation, proposing a relation between this situation and the local situation through a sub-ordered additive contribution:
\begin{equation}
\delta C_{ij} \sim exp[-N_c^{\frac{1}{3}} \Lambda^2_{QCD} (x_1-x_2)^2].
\end{equation}

In the local limit, Eq.(10) can give a prefactor which is of $O(1/\sqrt{N_c})$ for the localized tetraquark operator, without changing the leading-order's counting.\\

As one can see, for the local ``Type-B'' tetraquark operator, when we deal with the connected term of Eq.(8), this prefactor tends to turn down the overall counting of this term by $O(1/\sqrt{N_c})$. To maintain the counting, we have to raise the vertex order by $O(\sqrt{N_c})$, and this operation will give a vertex ordered $1/\sqrt{N_c}$. Therefore, by using this proposition, the counting of the connected term in Eq.(8) and the followed conclusions turns to be Weinberg's again.\\

While for nonlocal ``Type-A'', this proposition has no effect. Its vertex remains the same as $O(1/\sqrt{N_c})$. In this way, both ``Type-A'' and ``Type-B'' contribute the same decay width ordered $1/N_c$.\\

Combining above proposition and previous obtained counting similarities, we can get an even more important result. If we assign the prefactor $1/\sqrt{N_c}$ to tetraquark operator of ``Type-B'', then we can equally state that the creating or destroying order of such a tetraquark operator will be $O(1/\sqrt{N_c})$.\\

At the first sight, this seems to destroy our counting similarity between tetraquark and gluonic operators. However, since Eq.(10) is hand-added, we'd better discover its natural reason by insisting on counting similarity.\\

For convenience, we just replace the gluonic operators in Eq.(9) by tetraquark operators of ``Type-B'' as following:
\begin{equation}
<\mathcal{Q}_1 \cdots \mathcal{Q}_m \mathcal{B}_1 \cdots \mathcal{B}_n >_c.
\end{equation}

As a result of the change of the creating or destroying order of such tetraquark operators, Eq.(11) now counts like $O(N_c^{1-N/2})$ with $N=m+n$. And the corresponding decay width is of $O(N_c^{2-N})$. This reveals a mixing of tetraquark and mesonic operators in the counting.\\

This mixing also happens to ``Type-A'' situation, because in this situation, tetraquark operators are just similar with mesonic operators, and the corresponding counting is simply $O(N_c^{1-n/2})$. One can check Sec.III to verify this statement.\\

So, complying with Weinberg's result, together with the counting similarities, we find a mixing of operators in the counting. This mixing requires the hand-added term, i.e., Eq.(10). And such a mixing is consistent with our counting similarities naturally.

\section{Summary and Conclusions}
In this work, we first reviewed the preclusion of exotic meson states in large-$N_c$ limit QCD($\mathbf{F}$) briefly. Then we shortly reviewed the opposite proposal to such preclusion given by Weinberg, combining some followed viewpoints, we got the different orders for a tetraquark operator to create or destroy a normalized one-tetraquark state. By comparing the tetraquark operators with the mesonic/gluonic operators, we proposed that they are similar in the counting.\\

On the one hand, because the operators' creating or destroying orders are the key to get the vertex orders and the related decay widths of Eq.(8), and what we actually used are just such orders without the consideration of their actual physical properties, it's pretty legal to assert such similarities.\\

On the other hand, appropriate using of both similarities can simplify the analysis for tetraquark interactions (including decays, scatterings, etc.), just like what we have showed in Sec.III.\\

Furthermore, we can get an unified formula to decide the order of the tetraquark-including vertex (like the form of Eq.(8)). Remember the correlation of glueball-meson interactions and mixing processes, i.e., Eq.(9) which counts like $O(N_c^{1-m-n/2})$. Now that we know the similarities between tetraquark operators and mesonic/gluonic operators, we can directly use this correlation's counting to get the vertices of both ``Type-A'' and ``Type-B'': For ``Type-A'', $m=0,n=3$; while for ``Type-B'', $m=1,n=2$.\\

Meanwhile, our result about mixing property is almost the same as Ref.\cite{cohen}, although it starts from the two-index antisymmetric representation (QCD($\mathbf{AS}$)) \cite{corrigan}\cite{eb}. This also implies some new connection between these two representations, QCD($\mathbf{F}$) and QCD($\mathbf{AS}$), beyond the $N_c=3$ situation. This is worth to be investigated in the future.

\begin{center}
\textbf{Acknowledgement}
\end{center}

The authors appreciate R.F.Lebed and S.Peris so much for the helpful and inspirational discussions. This work has been supported by National Natural Science Foundation of China (No.11275017 and No.11173028).

\bibliographystyle{unsrt}

\begin{thebibliography}{99}

\bibitem{witten} E.Witten, {\it Nucl.Phys.B}, 160. 57. (1979). \par
\bibitem{thooft}  G.'t Hooft, {\it Nucl.Phys.B }, 72. 461. (1974). \par
\bibitem{coleman} S.R.Coleman, ``{\it Aspects of Symmetry} '', Cambridge University Press, Cambridge, England, (1985), pp.373-378. \par
\bibitem{reports} B.Lucini and M.Panero, {\it Physics Reports}, 526. 93.(2013); B.Lucini and M.Panero, {\it Progress in Particle and Nuclear Physics}, 75. 1.(2014)\par
\bibitem{weinberg2} S.Weinberg, {\it Phys.Rev.Lett}, 110. 261601. (2013). \par
\bibitem{lebed} R.F.Lebed, {\it Phys.Rev.D}, 88. 057901. (2013). \par
\bibitem{witten2} Exactly pp. 59-62 of Ref.[2]. \par
\bibitem{peris} M.Knecht and S.Peris, {\it Phys.Rev.D}, 88. 03616. (2013). \par
\bibitem{commu2} S.Peris (private communication). \par
\bibitem{commu} R.F.Lebed (private communication). \par
\bibitem{cohen} T.D.Cohen and R.F.Lebed, {\it Phys.Rev.D}, 89. 054018. (2014). \par
\bibitem{corrigan} E.Corrigan and P.Ramond, {\it Phys.Lett.B} 87. 73. (1979). \par
\bibitem{eb} E.B.Kiritsis and J.Papavassiliou, {\it Phys.Rev.D.} 42. 12. (1989).\par

\end{thebibliography}
\renewcommand{\bibname}{\zihao{3} \bf {References}}

\end{document}